\documentstyle[12pt]{article}
\def\setfonts{%
  \font\frbig=eufm10 scaled \magstep1 
  \font\frscr=eufm10
  \font\frscrscr=eufm8
  \newfam\frfam
  \textfont\frfam=\frbig
  \scriptfont\frfam=\frscr
  \scriptscriptfont\frfam=\frscrscr
  \def\fr{\fam\frfam}

  \font\openbig=msbm10 scaled\magstephalf
  \font\openscr=msbm8 
  \font\openscrscr=msbm8
  \newfam\openfam
  \textfont\openfam=\openbig
  \scriptfont\openfam=\openscr
  \scriptscriptfont\openfam=\openscrscr
  \def\open{\fam\openfam}
  }

\setfonts

\def\gs{{\fr s}}

\def\oR{{\open R}}
\def\oV{{\open V}}

\def\oN{{\open N}}

\begin{document}
\baselineskip=20pt

\begin{center}
{\Large\bf
 TEST FUNCTION SPACE FOR WICK POWER SERIES
}

\bigskip
{\large
A.~G.~Smirnov$^{_{1}}$, M.~A.~Soloviev}\footnote{
Lebedev Physics Institute, RAS, Moscow, Russia.
}
\end{center}

\vspace{0.3cm}

\begin{center}
{\large
Abstract}
\end{center}

\bigskip
We derive a criterion that is convenient for applications and exactly
characterizes the test function space on which the operator realization of
a given series of Wick powers of a free field is possible. The suggested
derivation does not use the assumption that the metric of the state space
is positive and can therefore be used in a gauge theory. It is based on the
systematic use of the analytic properties of the Hilbert majorant of the
indefinite metric and on the application of a suitable theorem on the
unconditional convergence of series of boundary values of analytic
functions.

 \section{ Introduction}

We consider convergence conditions for infinite series of the form
 \begin{equation}
  \sum_{k=0}^\infty d_k:\phi^k:(x)
 \label{lab1}
 \end{equation}
with respect to the Wick powers of a free field $\phi(x)$. Jaffe
\cite{Jaffe} first performed an analysis of this kind in determining the
Borchers class of a massive free field in the two-dimensional space-time.
He showed that the normally ordered entire functions of exponential type
for this field are well defined as operator-valued tempered distributions
satisfying all Wightman axioms. Subsequent works by Jaffe and some other
authors showed that Wick series that are ultradistributions and
hyperfunctions should also be included in the Borchers class and that it is
natural to consider even more singular generalized functions in extending
the axiomatic approach to nonlocal fields. Rieckers
\cite{Rieckers} showed that for any Wick series of a massive free field in
a space-time of dimension ${\tt d}\geq2$, there is a suitable
Gelfand-Shilov space $S^\beta$ such that this series is convergent under
averaging with test functions belonging to $S^\beta$. The generalization of
the technique in \cite{Jaffe,Rieckers}, which is essentially based on the
positive definiteness of the two-point function $\Delta^+(x;m)$, to fields
with an indefinite metric is far from obvious, although precisely fields
with an indefinite metric are of particular interest because the exact
solutions of some simple gauge models, including the Schwinger model, can
be expressed in terms of the normal exponentials of such fields. The
problem of an adequate choice of test functions when the positivity
condition is violated was raised by Wightman \cite{Wightman} and was
analyzed in \cite{Pierotti, MoschStroc, Moschella} for the examples of a
massless scalar two-dimensional field and a dipole ghost field. An
important observation in \cite{Pierotti, MoschStroc, Moschella} is that the
operator realization of the field (\ref{lab1}) in the case of an indefinite
metric is only possible under some conditions on the test functions that
are more restrictive than those under which the Wightman functions of this
field, which are calculated using the Wick theorem, are defined. The direct
analysis of series convergence in the momentum representation by analogy
with Rieckers' analysis is extremely laborious. In \cite{Soloviev1}, a
different approach was suggested based on using the analytic properties of
the Hilbert majorant of the indefinite metric in the coordinate
representation and investigating the convergence (in an appropriate
topology) of series of analytic functions instead of series of
distributions that serve as the boundary values of these functions. We show
that a criterion can be thus elaborated that solves the problem in the
general form and permits easily finding the test function space for any
Wick series (\ref{lab1}) with given coefficients proceeding from the
infrared and ultraviolet behavior of the two-point function of the free
field $\phi(x)$.

We consider the generalized Gelfand-Shilov spaces
$S^b_a$ and $S^b$ \cite{GelfandShilov} and use the results in
\cite{Soloviev2} that permit operating with functionals in the dual spaces
$S^{\prime b}_a$ and $S^{\prime b}$ as simply as with the usual Schwartz
distributions. The definitions needed for the proofs are given below. Here,
we only mention that the indices $a$ and $b$, which were number sequences
in the original statement in \cite{GelfandShilov}, can now be identified
with the corresponding indicator functions characterizing the decrease at
infinity of the test functions in $S^b_a$ and of their Fourier transforms
and that the elements of $S^b$ decrease faster than any power of
$1/|x|$ as $|x|\to\infty$. The elements of the most frequently used spaces
$S^\beta_\alpha$ in this scale have an exponential decrease of the order
$1/\alpha$ in the coordinate representation and of the order $1/\beta$ in
the momentum representation.

In Sec. 2, we present the necessary facts related to quantization with an
indefinite metric and specify the precise meaning of the convergence of the
series (\ref{lab1}) in the pseudo-Hilbert state space. We prove the key
convergence theorem for series of boundary values of analytic functions in
Sec. 4. An extension lemma for functionals that permits proving this
theorem rather simply is derived in Sec. 3. We obtain the ultimate
general convergence criterion for the Wick series of a free field in Sec.
5. In Sec. 6, we give applications to the case of a positive metric that
strengthen the results in \cite{Rieckers}. We consider an example of entire
functions of the two-dimensional massless free field that have a finite
order of growth in Sec. 7. In this example, the singularities are
logarithmic, and the use of the generalized spaces $S^b_a$ instead of
$S^\beta_\alpha$ permits characterizing the related test function space
much more precisely. Concluding remarks are presented in Sec. 8.

 \section{Graph limits of Wick power series}

Under the positivity condition, the operator realization of a free field
$\phi(x)$ is reconstructed from its two-point function
$w(x-x')$ up to a unitary equivalence. The major difficulty in generalizing
the reconstruction theorem to the indefinite-metric case lies in properly
relating to $w$ a distribution $w_{{\rm maj}}(x,x')$ of a positive type
that determines the Hilbert topology with respect to which the completion
gives the state space ${\cal H}$. The solution of this problem is described
in \cite{MorchioStrocchi, BLOT}, and we take the resulting mathematical
structure as the starting point. The space ${\cal H}$ is endowed by two
sesquilinear forms $\langle\cdot,\cdot \rangle$ and $(\cdot,\cdot)$ that
are determined by $w$ and $w_{{\rm maj}}$ and are related by the formula
\begin{equation}
\langle\Phi,\Psi \rangle=(\Phi,\theta\Psi), \qquad \Phi,\Psi\in {\cal H},
\label{lab2}
\end{equation}
where
 $\theta$ is an involutory and Hermitian operator with respect to the
positive form $(\cdot,\cdot)$ (it is known as the Krein operator).
This pseudo-Hilbert structure is compatible with the Fock structure
by which the space
${\cal H}$ is also endowed because we deal with a free theory.
The brackets $\langle\cdot,\cdot \rangle$ and $(\cdot,\cdot)$ are
expressed in terms of $w$ and $w_{{\rm maj}}$ at the one-particle level by
the formulas
$$
\langle\phi(f)\Psi_0,\phi(g)\Psi_0 \rangle=\int \bar f(x)w(x-x')g(x')\, {\rm
d}x {\rm d}x' ,
$$

\begin{equation}
(\phi(f)\Psi_0,\phi(g)\Psi_0)=\int \bar
f(x)w_{{\rm maj}}(x,x')g(x')\, {\rm d}x {\rm d}x' ,
\label{lab4}
\end{equation}
where
 $\Psi_0$ is the vacuum state and $f$, $g$
are test functions belonging to the Schwartz space. The compatibility with
the Fock structure means that the relations
$$
 \langle:\phi(f_1)\ldots\phi(f_m):\Psi_0,\,
:\phi(g_1)\ldots\phi(g_n):\Psi_0\rangle
 = \delta_{mn} \sum_\pi \prod_j
 \langle\phi(f_j)\Psi_0,\,\phi(g_{\pi(j)})\Psi_0\rangle
$$
\begin{equation}
 (:\phi(f_1)\ldots\phi(f_m):\Psi_0,\,
:\phi(g_1)\ldots\phi(g_n):\Psi_0)
 = \delta_{mn} \sum_\pi \prod_j
 (\phi(f_j)\Psi_0,\,\phi(g_{\pi(j)})\Psi_0),
\label{lab6}
\end{equation}
hold, where the summation extends over all permutations of indices.

In what follows, we use the properties that the majorant
$w_{{\rm maj}}$ must have by its sense and that can be easily verified
in every specific case. Namely, we assume that $w_{{\rm maj}}$ as well as
$w$ belongs to the class $S'$ of tempered distributions and inherits the
spectral properties of $w$ in the sense that the function
$$\hat w_{{\rm
maj}}(p,p')=\int e^{ipx+ip'x'} w_{{\rm maj}}(x,x')\,{\rm d}x {\rm d}x'$$
satisfies the condition
$$
{\rm supp}\, \hat  w_{{\rm maj}}(p,p') \subset
\bar \oV_+\times \bar \oV_-,
$$
 where $\oV_+$ and $\oV_-$ are the upper and lower light cones
and the bar symbolizes the closure. We stress that the majorant need not be
translation invariant, and we therefore do not impose the condition
$p+p'=0$. In terms of the coordinate representation, the above assumption
means that $w_{{\rm maj}}$ is the boundary value of an analytic
function ${\bf w}_{{\rm maj}}(z, z')$ that is holomorphic in the
tubular domain $\{z,z':\, y= {\rm Im\,}z\in \oV_-,\, y'= {\rm Im\,}z'\in
\oV_+\}$ and satisfies the condition
\begin{equation}
 |{\bf w}_{{\rm maj}}(z, z')|\le
 C(V)\,(1+|z|+|z'|)^\omega\, (|y|+|y'|)^{-\sigma}\quad
 (y,\, y'\in V),
\label{lab8}
\end{equation}
where $V$ is an arbitrary compact \footnote{The compactness means that
the inclusion $\bar V\setminus\{0\}\subset \oV_-\times \oV_+$ holds.
Here and henceforth, the norm $|\cdot|$ is assumed to be Euclidean.}
cone in
$\oV_-\times \oV_+$ and the numbers $\omega\geq 0$ and
$\sigma\geq 0$ do not depend on $V$.

The convergence of the series (\ref{lab1}) as well as the existence of the
Wick monomials themselves, which are formally defined by the
relation $$ :\phi^k:(x) = \lim_{x_1,\ldots, x_k \rightarrow
   x}:\phi(x_1)\ldots\phi(x_k):\,,
$$
should be established in the topology defined by the scalar
product $(\cdot,\cdot)$. A more exact statement of the definition is the
following. Let
$f\in S(\oR^{\tt d})$, where ${\tt d}$ is the space-time dimension. We take
an arbitrary $\delta$-shaped sequence $\delta_\nu(x)$ (i.e., convergent to
the Dirac delta function in $S'$) of nonnegative test functions that also
belong to $S(\oR^{\tt d})$ and consider the sequence of operators
\begin{equation}
\varphi_\nu(k,f)=\int
:\phi(x_1)\ldots\phi(x_k):  f(x_1)\prod_{j=2}^k \delta_\nu(x_1-x_j)\,{\rm
 d}x_1\ldots {\rm d}x_k
\label{lab9}
\end{equation}
with the domain $D_0$ composed of linear combinations of
  $\Psi_0$ and vectors of the form
  $$
 \Psi= \int
 :\phi(x_1)\ldots\phi(x_n):g(x_1,\ldots, x_n)\,{\rm d}x_1\ldots {\rm d}x_n
 \,\Psi_0,\qquad g\in S(\oR^{n\,\!{\tt d}}),\quad n=1,2,\ldots.
   $$
  It can be shown that the sequence (\ref{lab9}) has a strong graph limit
not depending on $\delta_\nu(x)$, which is precisely the limit denoted by
  $:\!\phi^k\!:\!(f)$.
Its domain contains all vectors of the form
\begin{equation}
\prod_{j=1}^n:\phi^{k_j}:(g_j)\Psi_0,
\label{lab10}
\end{equation}
where $g_j\in S(\oR^{\tt d})$,
whose linear span serves as the invariant domain of all monomials.
Each monomial is a tempered operator distribution satisfying the Lorentz
covariance, locality, and spectral conditions. It should be noted that the
technique below permits substantially simplifying the routine derivation of
these facts, but we do not dwell on this point.

We now introduce the notation
\begin{equation}
\varphi_N(f)=\sum_{k\le N} d_k:\phi^k:(f)
\label{lab11}
\end{equation}
and analyze the convergence conditions for the series (\ref{lab1}).
We first consider the action of the partial sums in
(\ref{lab11}) on the vacuum state. According to the definition of
$:\phi^k:$ and formulas (\ref{lab4}) and (\ref{lab6}), we have
\begin{equation}
 \|\varphi_N(f)\Psi_0\|^2=\sum_{k\le N} |d_k|^2\,\|\!:\phi^k:(f)\Psi_0\|^2=
\sum_{k\le N} k!\,|d_k|^2\,w_{{\rm maj}}^k(\bar f\otimes f),
\label{lab12}
\end{equation}
whence the effect of the absence of positivity is clearly seen, namely, the
series in the right-hand side has the same structure as the series of
distributions representing the two-point function of the field
(\ref{lab1}) with
$w$ replaced by $w_{{\rm maj}}$.
For the operator realization of the field (\ref{lab1}), its arbitrary
repeated action on $\Psi_0$ must be defined.
We indicate the domain of definition in
${\cal H}$ that is invariant not only with respect to this action but also
with respect to the action of any field determined by a series with
coefficients
$d_k^{\prime}$ such that $|d_k^{\prime}|\le C\, |d_k|$, where $C$ is a
constant depending on the series.
This condition establishes a preorder on the set of the series. We let
$\triangleleft$ denote this preorder and say that the series with the
coefficients $d_k^{\prime}$ is subordinate to the series with the
coefficients~$d_k$.

{\bf Lemma 1.} {\it
Let $\phi$ be a neutral scalar free field acting in the
pseudo-Hilbert space ${\cal H}$, let $\gs (\phi)$ be the
series $(\ref{lab1})$ in the Wick powers of this field with $d_1\ne 0$, let
$E$ be a barrelled space of test functions that is dense in $S$, and let
$D^E$ denote the subspace of ${\cal H}$ generated by $\Psi_0$ and by the
vectors of the form $(\ref{lab10})$, where $g_j\in E$. If for every
positive integer $n$ and for arbitrary sets of series $\gs^{(j)}
\triangleleft \gs$ and functions $f_j\in E$, $1\le j\le n$,
the sequence of vectors
\begin{equation}
\Phi_N=\sum_{k_j\le N\atop 1\le j\le n}
d_{k_1}^{(1)}\ldots d_{k_n}^{(n)} :\phi^{k_1}:(f_1)\ldots :\phi^{k_n}:(f_n)
\Psi_0,\quad N=1,2,\ldots ,
\label{lab13}
\end{equation}
where $d_{k_j}^{(j)}$ are the coefficients of $\gs^{(j)}$, converges in the
norm of ${\cal H}$, then the series $\gs(\phi)$ with the domain $D^E$
converges in the sense of the strong graph limit to a field $\varphi$
that is an operator-valued generalized function over $E$. Moreover, any
Wick series subordinate to $\gs$ is convergent, and all these
fields have the common domain $D^E(\gs)$, which is dense and invariant.}

{\bf Proof}.  Let
$\Psi_N$, $\Psi'_N\in  D^E$, $\Psi_N\to \Psi$, $\Psi'_N\to \Psi$,
$\varphi_N(f)\Psi_N\to \Phi$, and $\varphi_N(f)\Psi'_N\to \Phi'$.
By the definition of the strong graph limit \cite{ReedSimon},
we should show that $\Phi=\Phi'$. Let
   \begin{equation}
 \Omega=:\phi(f_1)\ldots\phi(f_n): \,\Psi_0,\qquad f_j\in E,\quad 1\leq
 j\leq n.
   \label{lab14}
   \end{equation}
It follows from the Wick theorem that
$$
\varphi_N(f)^*|D^E=\sum_{k\le N}\, \bar
   d_k\,:\phi^k:(\bar f),
$$
where the superscript $*$ symbolizes the pseudo-Hermitian conjugation.
Since $d_1\ne 0$, the sequence $\varphi_N(f)^*\,\Omega$ is a finite linear
combination of sequences (\ref{lab13}) of particular form and consequently
is convergent. Hence,
  $$ \langle
  \Phi-\Phi',\,\Omega\rangle=\lim_{N\to\infty} \langle
 \varphi_N(f)(\Psi_N-\Psi'_N),\,\Omega\rangle=\lim_{N\to\infty}
 \langle \Psi_N-\Psi'_N,\,\varphi_N(f)^*\,\Omega\rangle=0,
   $$
and therefore the desired relation
$\Phi=\Phi'$ holds because the linear span of the vectors of the form
 (\ref{lab14}) is dense in ${\cal H}$, which follows from the density of
$E$ in $S$. To show this, suppose
$\bar \Omega$ is a vector of the same form but with the test functions
$g_j\in S$ and consider $f_{j\,\nu}\in E$,
$f_{j\,\nu}\to g_j$ in $S$. We write $g=g_1\otimes\dots\otimes g_n$,
 $f_\nu=f_{1\nu}\otimes\dots\otimes f_{n\nu}$
and $\Omega_\nu=:\phi(f_{1\nu})\dots \phi(f_{n\nu}):\Psi_0$. Then
 $$
 \|\bar \Omega-\Omega_\nu\|^2=
T_n\left((\overline{f-f_\nu})\otimes(f-f_\nu)\right),
 $$
where
$$
T_n(x,x')=\sum_\pi \prod_{j=1}^n
w_{{\rm maj}}(x_j,x'_{\pi(j)}) .
$$
according to (\ref{lab4}) and (\ref{lab6}). The membership relation
$T_n\in S^\prime$ implies $\Omega_\nu\to \Omega$,
and the abovementioned denseness property holds by virtue of the cyclicity
of the vacuum.

Let $\varphi(f)$ denote the graph limit of the sequence (\ref{lab11}).
A pair $\{\Psi,\Phi\}$ belongs to the graph of this operator
if there is a sequence $\Psi_N\in D^E$ such that
 $\Psi_N\to\Psi$ and $\varphi_N(f)\Psi_N\to \Phi$. The partial sums of any
subordinate series $\gs^{(j)}(\phi)$ also have limits. The vectors
$\prod_{1\le j\le n}\varphi^{(j)}(f_j)\Psi_0$, where $f_j\in E$ and
  $n=1,2,\ldots$, are well defined, and the linear span of
$\Psi_0$ and of all vectors of this form can be taken as the subspace
 $D^E(\gs)$. This subspace is invariant and dense in ${\cal H}$. The
role of the barrelledness condition for $E$ is that for
barrelled spaces, the uniform boundedness principle holds and
the pointwise convergence of the sequence of continuous linear
mappings $f\to \varphi_N(f)\Psi_N$ from $E$ to ${\cal H}$ implies the
continuity of the limit mapping $f\to \varphi(f)\Psi$ (see
theorem III.4.6 in \cite{Schaefer}). As a consequence,
$\langle\Phi,\,\varphi(f)\Psi\rangle$ is a generalized function over the
space $E$ for any $\Phi,\Psi\in D^E(\gs)$.  The lemma is proved.

The expression for the $n$-point vacuum expectation value of the field
$\varphi$ given by the Wick theorem is a power series in
$n(n-1)/2$ variables $w(x_j-x_m)$ and can be written as
   \begin{equation}
   \langle
     \Psi_0,\, \varphi(x_1)\ldots \varphi(x_n)\Psi_0\rangle
     =\sum_KD_K\,W^K\,,
   \label{lab15}
   \end{equation}
where $K$ is an integer-valued vector with nonnegative components
  $k_{jm}$, $1\leq j<m\leq n$, and
$$
   W^K\stackrel{\rm
  def}{=}\prod_{j<m}w(x_j-x_m)^{k_{jm}}\,.
$$
The combinatorial analysis related to the Wick theorem shows
\cite{Jaffe} that
 $$
  D_K={\kappa!\over K!}\prod_{1\leq j\leq n}d_{\kappa_j},
 $$
 where  $\kappa_j=k_{1j}+\ldots+k_{j-1,j}+k_{j,j+1}+\ldots+k_{jn}$ is
the total number of pairings in the given term of the series that involve
the argument
 $x_j$, and we follow the usual convention
$$
K!= \prod_{j<m}k_{jm}!\,,\quad
\kappa!= \prod_{1\leq j\leq n}\kappa_j!\,.
$$

We use one more simple lemma to derive specific convergence conditions for
the multi-index series whose partial sum is the vector (\ref{lab13}).

    {\bf Lemma 2.} {\it
Let $I$ be a countable set of indices and let $\{\Phi_k\}_{k\in  I}$ be a
family of elements in the Hilbert space with the scalar product
$(\cdot,\cdot)$. If the double number series $\sum_{k,l\in
I}(\Phi_k,\,\Phi_{l})$ is absolutely convergent, then the
family $\{\Phi_k\}_{k\in I}$ is unconditionally summable.}

{\bf Proof.} By definition, unconditional summability means that for any
one-to-one mapping
$\nu\to k_\nu$ of the set of nonnegative integers
${\oN}$ onto $I$, the series $\sum _{\nu\in {\oN}} \Phi_{k_\nu}$
is convergent, i.e., for any $\epsilon >0$ there is an index
$\lambda_0$ such that $\|\sum_{\lambda\leq\nu\leq\Lambda}
\Phi_{k_\nu}\|< \epsilon$ for $\lambda_0\leq \lambda \leq\Lambda$.
Since the square of the norm involved does not exceed
$\sum_{\lambda\leq\mu,\nu\leq\Lambda} |(\Phi_{k_\mu},\,\Phi_{k_\nu})|$,
the assertion of the lemma follows from the elementary fact that
a convergent series of positive numbers remains convergent and satisfies
the Cauchy criterion for convergence under an arbitrary permutation of its
terms. Lemma 2 is proved.

Thus, the assumptions of Lemma 1 hold if the number series
$$
  \sum_{k_1,\ldots, k_n\atop l_1,\ldots, l_n}
  (\prod_{1\le j\le n}d_{k_j}:\phi^{k_j}:(f_j)\Psi_0,
  \,\prod_{1\le j\le n}d_{l_j} :\phi^{l_j}:(f_j)\Psi_0),
$$
which represents $\|\prod_{1\le j\le n}\varphi(f_j)\Psi_0\|^2$
in this case, is absolutely convergent for some $n$ and for all $f_j\in E$.
In what follows, we keep to the argument indexing that corresponds to the
formal representation of this expression in the form
  $$
 \int (\varphi(x_1)\ldots\varphi(x_n)\Psi_0,\,
  \varphi(x_{n+1})\ldots\varphi(x_{2n})\Psi_0)\overline{f_1(x_1)\ldots
  f_n(x_n)}f_1(x_{n+1})\ldots f_n(x_{2n})\,{\rm d}x_1\ldots{\rm d}x_{2n}.
  $$
Bringing each of the products of Wick powers to the totally normally
ordered form
(see Appendix A in \cite{Jaffe}) and then applying (\ref{lab6}), we obtain
\begin{equation}
    \|\prod_{1\le j\le
    n}\varphi(f_j)\Psi_0\|^2=\sum_KD_K\,W^K(\bar f\otimes f)\,,
   \label{lab18}
   \end{equation}
where $f=f_1\otimes\ldots\otimes
f_n$. This expression is similar to the formula (\ref{lab15}) with the
obvious change $n\to 2n$ in the definition of multi-indices $K$ and $\kappa$.
Moreover,
$$
D_K=(\kappa!/K!)\prod_{1\leq j\leq n}\bar d_{\kappa_j}\prod_{n\leq
j\leq 2n}d_{\kappa_j}.
$$
A more important feature is that in this case we have
\begin{equation}
   W^K=\prod_{1\le j<m\le n}w(x_m-x_j)^{k_{jm}}\prod_{ n+1\le j<m\le 2n}
   w(x_j-x_m)^{k_{jm}}\prod_{1\le j\le n\atop n+1\le m\le 2n}w_{{\rm
  maj}}(x_j,x_m)^{k_{jm}}.
   \label{lab19}
   \end{equation}
The problem has now been reduced to finding convergence conditions for a
series of distributions of a special structure whose terms serve as
boundary values of analytic functions in the same tubular domain. The major
difficulty lies in determining the type of the singularity of the sum of
the series if it is summable. The singularity of a functional is related to
the behavior of its Fourier transform at infinity, and we characterize it
by estimating the convolutions of the Fourier transform with rapidly
decreasing test functions.

     \section{Extension lemma}

    {\bf Lemma 3.} {\it Let $u\in S^{\prime a}_{b_0}$ and $b(s)\leq
    b_0(s)$. If the convolutions of a functional $u$ with the functions
$g\in S^{a,A}_{b_0,B}$ satisfy the inequality
при некотором $B>0$ и любом $A>0$ справедлива оценка
\begin{equation}
    |(u*g)(p)|\le C_{\epsilon,A}\,\|g\|_{A,B}\, b(\epsilon |p|)
   \label{lab20}
   \end{equation}
with an arbitrary $\epsilon>0$ for some $B>0$ and any $A>0$, then the
functional $u$ has a unique continuous extension to the space $S^a_b$.
If the family of functionals in $S^{\prime a}_{b_0}$ satisfies the uniform
estimate $(\ref{lab20})$, then the extended functionals form a bounded set
in $S^{\prime a}_b$.  Similarly, if $u\in S^\prime_{b_0}$ and the
inequality
\begin{equation}
|(u*g)(p)|\le
     C_{\epsilon}\|g\|_{B,N}\, b(\epsilon |p|)
   \label{lab21}
   \end{equation}
with an arbitrary $\epsilon>0$ and some $N$ $($which can depend on
$\epsilon$$)$ holds for some $B$ and all $g\in S_{b_0,B}$, then the
functional $u$ has a unique extension to an element of the space
$S'_b$, and extending the family of functionals results in a bounded set in
the case of the uniform estimate.}

Lemmas of this type were used in \cite{Soloviev2}, but here we give this
refined statement suitable for our further aims. We stress that the
indicator functions of these spaces possess some regularity properties
according to their definition. In particular, they are convex and
monotonically increasing beginning with unity, and therefore
\begin{equation}
      b((s_1+s_2)/2)\leq
     (b(s_1)+b(s_2))/2\leq b(s_1)b(s_2).
   \label{lab22}
   \end{equation}
Moreover, the inequality
\begin{equation}
   sb(s)\le Cb(\lambda s)
   \label{lab23}
   \end{equation}
with some constants $C$ and $\lambda$ must hold, which ensures the
existence of the operation of multiplication by the independent variable.
The two properties expressed by (\ref{lab22}) and (\ref{lab23})
are essential for deriving Lemma 3, and the sequence
$b_0$ must even satisfy the condition
\begin{equation}
   b_0^2(s)\leq C b_0(\lambda s)\,,
   \label{lab24}
   \end{equation}
which is stronger than (\ref{lab23}). Here, we only prove Lemma 3 for the
case of functionals of the class $S'_{b_0}$, which was not considered in
\cite{Soloviev2}. We recall that $S_{b_0}$ is the union of the countably
normed spaces $S_{b_0,B}$ and that the norm
$\|g\|_{B,N}$ in (\ref{lab21}) has the form
$$
\|g\|_{B,N} =
         \sup_{q}\max_{|\kappa|\le  N}
         |\partial^{\,\kappa}g(q)|\,b_0(|q|/B).
$$
The desired extension $\tilde u$ can be defined by the formula
   \begin{equation}
   (\tilde u,
h)=\int(u, h_0(p-\cdot)h(\cdot)) \,{\rm d}p,\quad h\in S_b,
   \label{lab25}
   \end{equation}
where $h_0$ is an arbitrary element in $S_{b_0,B/\lambda}$ with the
property
$$
\int h_0(p){\rm d}p=1.
$$
The formula (\ref{lab21}) implies the inequality
\begin{equation}
|(u,h_0(p-\cdot)h(\cdot))|\le C_\epsilon\, \|g_p\|_{B,N}\, b(\epsilon |p|),
   \label{lab26}
   \end{equation}
where $g_p(q)\stackrel{\rm def}{=} h_0(q)h(p-q)$. If $h\in S_{b,B_1}$,
where $B_1\geq B$, then applying the Leibniz formula gives
\begin{equation}
      \|g_p\|_{B,N} \leq 2^N
      \|h_0\|_{B/\lambda,N}\,\|h\|_{B_1,N}\,\sup_q
      \frac{b_0(|q|/B)}{b_0(\lambda|q|/B)b(|p-q|/B_1)}
   \label{lab27}
   \end{equation}
Consecutively applying the property (\ref{lab24}), the inequality $b(s)\leq
b_0(s)$ and the formula (\ref{lab22}) that the fraction in the right-hand
side of (\ref{lab27}) can be estimated from above by the
function $C/b(|p|/2B_1)$. Substituting it in (\ref{lab26})
and taking (\ref{lab23}) into account, we see that the integrand in
(\ref{lab25}) decreases faster than polynomially as
$|p|\to\infty$ and the integral specifies a linear functional on
$S_b$, whose continuity is guaranteed by the presence of the factor
$\|h\|_{B_1,N}$ in the right-hand side of (\ref{lab27}).
The constructed functional coincides with the original one on the functions
$h\in S_{b_0}$ because in this case, the integral (\ref{lab25})
exists not only for $u$ but also for any distribution in $S'_{b_0}$.
Consequently, the sequence of Riemann integral sums for
$\int h_0(p-\cdot)h(\cdot) \,{\rm d}p$ is weakly Cauchy in the space
$S_{b_0}$ and because this is a Montel space\footnote{In
\cite{GelfandShilov}, the term ``perfect space'' was used instead of
``Montel space'', which is now commonly accepted. See \cite{Schaefer}
for the properties of topological vector spaces we use here.}, it converges
in $S_{b_0}$ to an element that can only be $h$. This consideration also
proves that $S_{b_0}$ is dense in $S_b$ and the extension is hence unique.
If a family of functionals belonging to $S^\prime_{b_0}$ satisfies the
uniform estimate $(\ref{lab21})$, then the set of their extensions to
$S'_b$ is obviously weakly bounded, and because $S_b$ is a barrelled space,
this set is also strongly bounded.

      {\raggedright
      \section{Convergence conditions for series of boundary values of
analytic functions}}
As already mentioned, the distributions entering the multi-index series
(\ref{lab18}) are the boundary values of the holomorphic functions
${\bf v}_K$ in the same tubular domain. The base this domain is a convex
cone $V$, and for any compact subset $Q\subset V$, the condition of
polynomial boundedness
\begin{equation}
        |{\bf v}_K(x+iy)|\leq
       C(1+|x|)^\omega \qquad (y\in Q),
   \label{lab28}
   \end{equation}
holds, where the constants $C$ and $\omega$ generally depend on $Q$.
We first consider the series (\ref{lab18}) on the spaces $S^1_a$, where
the index $1$ means that $\beta =1$, i.e., the indicator function
$b(s)$ of these spaces is $e^s$. The dual spaces of $S^1_a$ consist of
hyperfunctions with the order of growth
$\sim a(|x|)$ at infinity.
As is known, hyperfunctions provide the widest framework for
constructing a local field theory. By definition, we
       have $S^1_a=\bigcup_{A,B>0} S^{1,B}_{a,A}$, where $S^{1,B}_{a,A}$ is
the Banach space of functions which are analytic in the domain $T_B=\{x+iy:
       |y|<1/B\}$, continuous in its closure and have the finite norm
\begin{equation}
\|f\|_{A,B}=\sup_{z\in T_B} |f(z)|a(|x|/A).
\label{lab29}
\end{equation}
In particular, these test functions have a special role in the problem
under study because the boundary value $v_K$
of an analytic function ${\bf v}_K$ on them can be represented in the form
\begin{equation}
        v_K(f)=\int{\bf v}_K(x+iy)f(x+iy)\,{\rm d}x\qquad
       (y\in V,\,\,|y|<1/B)\,,
\label{lab30}
\end{equation}
which is convenient for analysis.
Indeed, by the Cauchy-Poincar\'e theorem, the integral in the right-hand
side does not depend on $y$ and coincides with the left-hand side
by virtue of the relation
$\lim_{y\to 0}f(x+iy)=f(x)$, which holds in the topology of $S^1_a$ and of
course in that of $S$.

     {\bf Theorem 1.} {\it Let $V$ be an open convex cone, let
     $(v_K)_{K\in I}$ be a countable family of tempered distributions
     serving as boundary values of functions ${\bf v}_K(z)$
     holomorphic in $T^V= \{z:  {\rm Im}\,z \in V\}$ and
     satisfying $(\ref{lab28})$, and let $a$ and $b$ be the indicator
     functions of the space $S^b_a$ with the nontrivial subspace $S^1_a$.
     If there is a vector $\eta \in V$ such that
\begin{equation}
   \sum_{K\in I} \inf_{0<t<\delta} e^{st}\int\frac{|{\bf
    v}_K(x+it\eta)|}{a(|x|/A)}\, {\rm d}x\leq C_{\delta,\epsilon,A}
    b(\epsilon s)
   \label{lab31}
   \end{equation}
     for any positive $\delta$, $\epsilon$ and $A$, then the family
     $(v_K)_{K\in I}$ is unconditionally summable in $S^{\prime b}_a$.}

 {\bf Proof.}
 Let $f\in S^{1,B}_{a,A}$ and $|\eta|=1$. Consecutively applying
   the representation (\ref{lab30}), the definition (\ref{lab29}),
   and the inequalities (\ref{lab31}) with $s=0$ and $\delta<1/B$
   yields
$$
 \sum_{K\in I}|v_{K}(f)|\leq \|f\|_{A,B}\sum_{K\in I}
 \inf_{0<t<\delta} \int\frac{|{\bf
    v}_{K}(x+it\eta)|}{a(|x|/A)}\, {\rm d}x \leq C_{A,B}.
$$
 Therefore, the family of distributions $v_K$ is absolutely summable on
 each element of $S^1_a$.
Because this is a Montel space, its strong dual is also a Montel space,
and it contains an element $v$ to which the family in question is
unconditionally summable in the strong topology. We investigate the
convolution of the Fourier transform
$\hat v=u\in S^{\prime a}_1$ and a test function $g\in
S^{a,A}_{1,B}$ using the relation
\begin{equation}
 (u*g)(p)=\sum_{K\in I}\int {\bf v}_{K}
 (x+iy)e^{ip(x+iy)}f(x+iy)\,{\rm d}x,
   \label{lab33}
   \end{equation}
which holds for any $y$ belonging to both the domain $V$ and the base of
the analyticity domain of the function
$$
f(z)=\int g(-p)e^{ipz}\,{\rm d}p.
$$
The Fourier operator realizes a one-to-one bicontinuous mapping of
$S^a_1$ onto $S_a^1$, and there exist $A'> A$ and $B'> B$ such that
$\|f\|_{S^{1,B'}_{a,A'}}\le C\,\|g\|_{S^{a,A}_{1,B}}$. Therefore,
(\ref{lab31}) and (\ref{lab33}) imply the estimate
\begin{equation}
 |(u*g)(p)|\le C\,\|g\|_{A,B} \sum_{K\in I}\inf_{0<t<1/B'}\,e^{|p|t}\,
 \int\frac{|{\bf v}_{K}(x+it\eta)|}{a(|x|/A')}\,{\rm d}x \leq C_{\epsilon,
    A}\,\|g\|_{A,B}b(\epsilon |p|)\,,
\label{lab36}
\end{equation}
whence  $u\in S^{\prime  a}_b$ by Lemma 3.

Now let $f\in S^b_a$, $f_\mu \in S^1_a$, and $f_\mu\to
f$ in $S^b_a$. We fix a one-to-one mapping $\oN\to I:\,\nu\to K_\nu$
and write
$$
v_\lambda(f)=\sum_{\nu\leq \lambda} v_{K_\nu}(f)
$$
and
$$
   v(f)-v_\lambda(f)=v(f-f_\mu)+[v(f_\mu)-
   v_\lambda(f_\mu)]+v_\lambda(f_\mu-f)
$$
The Fourier transform of each $v_\lambda$ satisfies the inequality of the
form (\ref{lab36}) with the same constant $C_{\epsilon, A}$ for all
$\lambda$. Consequently, by Lemma 3, this sequence of distributions is
   bounded in $S^{\prime b}_a$. As in the case of an arbitrary reflexive
   space, the topology of $S^b_a$ coincides with the uniform convergence
   topology on bounded sets. The limiting relation
   $v_\lambda(f_\mu-f)\to 0$ therefore holds uniformly with respect to
   $\lambda$ as $\mu\to \infty$. For a fixed sufficiently large $\mu$ and
    an arbitrary $\varepsilon>0$, there is a number
    $\lambda_{\varepsilon,\mu}$ such that
    $|v(f_\mu)-v_\lambda(f_\mu)|<\varepsilon$ for
    $\lambda>\lambda_{\varepsilon,\mu}$. We therefore have $v_\lambda\to v$
    in the weak and consequently in the strong topologies of the Montel
    space $S^{\prime b}_a$. The theorem is proved.

Wick power series converging to nonlocal fields \cite{Rieckers} are of
 interest in field theory with positive metric. An analogue of Theorem 1 is
 useful in their analysis. The role of $S^1_a$ in its derivation is played
 by the space $S^0$ $(\beta=0)$, which is the Fourier transform of
 the space $S_0=C^\infty_0$ of smooth functions with compact support.
 More precisely, $S^0=\bigcup_{B>0}S^{0,B}$, where $S^{0,B}$ consists of
 the entire functions $f(z)$ such that the norm
$$
  \|f\|_{B,N}=\sup_{z=x+iy} (1+|z|)^N |f(z)|e^{-B|y|}
$$
is finite for any $N$.

    {\bf Theorem 2.} {\it As in Theorem $1$, let
   $(v_K)_{K\in I}$ be a countable family of tempered distributions
    serving as boundary values of the functions ${\bf v}_K(z)$
holomorphic in the domain $T^V$ and satisfying the condition
$(\ref{lab28})$. Let $b$ be the indicator function of the space
$S^b$. If there is a vector $\eta \in V$ such that
\begin{equation}
    \sum_{K\in I} \inf_{t>0} e^{st}\int\frac{|{\bf
    v}_K(x+it\eta)|}{(1+|x|)^N}\, {\rm d}x\leq C_\epsilon b(\epsilon s),
   \label{lab37}
   \end{equation}
for any $\epsilon>0$ and some $N(\epsilon)$, then the family
$(v_K)_{K\in I}$ is unconditionally summable in $S^{\prime b}$.}

{\bf Proof.} Let $f\in S^{0,B}$ and $|\eta|=1$.
 Then
$$
|f(x+it\eta)|\leq \|f\|_{B,N} e^{Bt}(1+|x|)^{-N}\,.
$$
In view of (\ref{lab30}) and (\ref{lab37}) with $s=B$, we conclude
that the family of distributions $v_K$ is absolutely summable on
every element of $S^0$. We let $u$ denote the Fourier transform of the
 limit and consider the convolution of $u$ with a function $g$ in
 $S_{0,B}$. By the definition of $S_{0,B}$, the support of $g$ lies in the
 ball $|p|<B$, and all the norms $\|g\|_{S_{0,B,N}}= \sup_{|p|\le
 B}\max_{|\kappa|\le N}|\partial^{\,\kappa}g(p)|$ are finite. Let
 $f(z)$ be defined as in the proof of Theorem 1. For a sufficiently large
 $B'$, we have a continuous embedding $S_{0,B}\to
 S^{0,B'}$, and for any $N'$ there exists an $N$ such that
 \begin{equation}
  \|f\|_{S^{0,B',N'}}\le C\,\|g\|_{S_{0,B,N}}.
 \label{lab1a}
  \end{equation}
Let $C_{\epsilon}$ and $N'(\epsilon)$ be the numbers involved in
(\ref{lab37}). Applying (\ref{lab33}), (\ref{lab1a}), (\ref{lab37}), and
(\ref{lab22}) consecutively, we obtain
$$
|(u*g)(p)| \leq
CC_{\epsilon}\,\|g\|_{B,N(\epsilon)} b(2B'\epsilon)b(2\epsilon |p|)\,.
$$
We now apply Lemma 3 to $u$ and to the family $u_{\lambda}$ of the Fourier
 transforms of partial sums $v_{\lambda}$, $(\lambda\in \oN)$, to
 conclude that $v\in S^{\prime b}$ and that the sequence
 $v_{\lambda}$ is bounded in $S^{\prime b}$. The remaining part of the
 proof coincides with the end of the proof of Theorem 1 because $S^b$
 are also Montel spaces.

  {\raggedright
      \section{Test function space for Wick series} }

When studying the behavior of analytic functions whose boundary
 values are distributions (\ref{lab19}), we can replace all functions ${\bf
 w}(z_j-z_m)$ with ${\bf w}_{{\rm maj}}(z_j,z_m)$ because
\begin{equation}
|{\bf w}(x-x'-2iy)|^2\leq |{\bf w}_{{\rm maj}}(x-iy,\, x+iy)|
\,|{\bf w}_{{\rm maj}}(x'-iy,\, x'+iy)|
\label{lab38}
\end{equation}
for all $y\in \oV_+$. Indeed, (\ref{lab2}) implies the inequality
$$
|\langle\phi(f)\Psi_0,\phi(g)\Psi_0 \rangle|\leq \|\phi(f)\Psi_0\|
\|\phi(g)\Psi_0\|\,.
$$
Taking $f(\xi)=(\nu/\sqrt\pi)^{\tt d}e^{-\nu^2(\xi-x-iy)^2}$ and
$g(\xi)=(\nu/\sqrt\pi)^{\tt d}e^{-\nu^2(\xi-x'-iy)^2}$ and writing the
left- and right-hand sides in this inequality as integrals over a
 plane in the analyticity domain and passing to the limit as
 $\nu\to\infty$, we immediately obtain
 (\ref{lab38}). Everywhere below, we assume that the imaginary parts of the
 first and second arguments of the function ${\bf w}_{{\rm maj}}$
 belong to the respective negative and positive half-lines $y_0$ and
 characterize its infrared and ultraviolet behavior with a pair of
 monotonic nonnegative functions
 $w_{{\scriptscriptstyle IR}}$ and $w_{{\scriptscriptstyle UV}}$.
 The first (second) function increases as the argument increases (as the
 argument decreases), and the inequality
\begin{equation}
      |{\bf w}_{{\rm maj}}(z, z')|\le C_0 +C_1\,w_{{\scriptscriptstyle
       IR}}(|z|+|z'|)+C_2\,w_{{\scriptscriptstyle UV}}(|y|+
       |y'|).
   \label{lab39}
   \end{equation}
 is assumed. By formula (\ref{lab8}), such characteristic functions
$w_{{\scriptscriptstyle IR}}$ and $w_{{\scriptscriptstyle UV}}$
exist and can be easily chosen in each specific case, particularly with
   preservation of the normalization condition $w_{{\scriptscriptstyle
   IR}}(0)=w_{{\scriptscriptstyle UV}}(\infty)=0$.

We seek the test function space for the series (\ref{lab1})
under some natural conditions on the coefficients of the series.
We assume that
\begin{equation}
d_k\geq0,\quad d_0=1,\quad
\lim_{k\to\infty} (k!d_{2k})^{1/k}=0.
   \label{lab40}
   \end{equation}
The first two conditions are purely technical and are justified because
every series is subordinate to a series satisfying these conditions.
In a theory with
a positive metric, the third condition ensures that the sum
of the series is a local field, and in the case of an indefinite metric,
it is necessary for the very existence of the sum.
Moreover, we assume that
\begin{equation}
d_kd_l \le Ch^{k+l} d_{k+l},
\label{lab41}
\end{equation}
where $C$ and $h$ are some constants. In this case,
the restrictions on the test functions under which the complete operator
      realization of the field (\ref{lab1}) is possible coincide
      essentially with those guaranteeing the convergence of this series
      on the vacuum vector.
If the conditions (\ref{lab40}) and (\ref{lab41}) hold, then
$d_k^2\leq Ch^{2k} d_{2k}$, and the two-point function $$
\langle\Psi_0,\varphi(x_1)\varphi(x_2)\Psi_0\rangle=
     \sum_kk!d_k^2 w(x_1-x_2)^k
$$
is analytic in a domain ordinary for the local theory because it is the
composition of $w$ and some entire function.
Similarly, the function $\sum_kk!d_k^2w_{{\rm maj}}(x_1,x_2)^k$ entering
(\ref{lab12}) inherits the analytic properties of $w_{{\rm maj}}$.

        {\bf Theorem 3.} {\it Let $\phi$ be a free field acting in the
pseudo-Hilbert space ${\cal H}$, and let the positive majorant of its
    correlation function satisfy the inequality $(\ref{lab39})$ in which
    $w_{{\scriptscriptstyle IR}}$ and $w_{{\scriptscriptstyle UV}}$
    are monotonic. Under the conditions
    $(\ref{lab40})$ and $(\ref{lab41})$ on the
    coefficients, the series $\varphi=\gs(\phi)$ in the Wick powers of this
    field and also any of its subordinate series are well defined
    as operator-valued generalized functions on every space $S^b_a$
    containing a nontrivial subspace $S^1_a$ whose indicator functions
    satisfy the inequalities
\begin{equation}
    \sum_k L^kk!d_{2k} w_{{\scriptscriptstyle IR}}(r)^k \le
      C_{L,\epsilon}\,a(\epsilon r),\quad \inf_{t>0}\,e^{s\tau}
      \sum_k L^kk! d_{2k} w_{{\scriptscriptstyle UV}}(t)^k \le
      C_{L,\epsilon}\, b(\epsilon s).
   \label{lab42}
   \end{equation}
for an arbitrarily large $L>0$ and an arbitrarily small $\epsilon>0$.}

{\bf Proof.} We set $v_K=D_K\,W^K$ and apply Theorem 1 to the series
    (\ref{lab18}).
    Let $e_0$ be the unit vector $(1,{\bf 0})$ in $\oR^{\tt d}$, let $\eta$
be a vector in $\oR^{2n{\tt d}}$ with the components
$\eta_j=-je_0$ for $1\leq j\leq n$ and $\eta_j=(j-n)e_0$ for $n+1\leq j\leq
2n$, and let $r=|x|$, where $x=(x_1,\ldots,x_{2n})$.
Applying (\ref{lab38}) and (\ref{lab39})
in view of the
monotonicity of the functions $w_{{\scriptscriptstyle  IR}}$ and
$w_{{\scriptscriptstyle UV}}$ yields
\begin{equation}
     |{\bf W}^K(x+it\eta)|\leq  3^{|K|-1}\left(C^{|K|}_0 +
     C^{|K|}_1\,w_{{\scriptscriptstyle IR}}(2r+2nt)^{|K|}+ C^{|K|}_2\,
    w_{{\scriptscriptstyle UV}}(t)^{|K|}\right).
   \label{lab43}
   \end{equation}
The condition (\ref{lab41}) and the inequalities $|K|!/K!\leq
(n(2n-1))^{|K|}$ and $\kappa!\leq|\kappa|!\leq 4^{|K|}(|K|!)^2$ following
from the well-known properties of polynomial coefficients imply
\begin{equation}
    D_K\leq C'h^{\prime |K|}|K|!d_{2|K|},
   \label{lab44}
   \end{equation}
where the constant $h'$ depends on $n$. The summation over the
multi-indices can now be replaced with that over the positive
integers. The number of multi-indices with a given norm
$|K|$ is equal to the binomial coefficient ${|K|+n(2n-1)-1\choose |K|}$.
This dependence on $|K|$ is polynomial and therefore insignificant.
Using (\ref{lab43}) and (\ref{lab44}), we obtain
$$
\sum_K|{\bf v}_K(x+it\eta)|\leq C''\left(\sum_{k=0}^\infty L^kk!d_{2k}
w_{{\scriptscriptstyle IR}}(2r+2nt)^k\right) \left(\sum_{k=0}^\infty
    L^kk!d_{2k}w_{{\scriptscriptstyle UV}}(t)^k\right).
$$
The first of the conditions (\ref{lab42}) and the analogue of
the inequality (\ref{lab22}) for the indicator function $a$ permits
majorizing the sum of the powers of $w_{{\scriptscriptstyle IR}}$
with the expression $C_{L,\epsilon}(t)a(4\epsilon r)$, where
$C_{L,\epsilon}(t)$ is a bounded function on every finite interval.
    Applying the condition (\ref{lab23}) to $a$ results in
    $(1+r)^Na(4\epsilon r)\leq C a(r/A)$ if an
    $\epsilon<1/(4A\lambda^N)$ is chosen. Taking $N=2n{\tt d}+1$, we obtain
    the estimate
$$
    \sum_K\int\frac{|{\bf v}_K(x+it\eta)|}{a(|x|/A)}\, {\rm
   d}x\leq C_{A,L}(t)\sum_{k=0}^\infty L^kk!d_{2k}w_{{\scriptscriptstyle
   UV}}(t)^k.
$$
Because $w_{{\scriptscriptstyle UV}}$ is monotonically increasing with
a decreasing argument, the infimum in the second formula in
(\ref{lab42}) occurs on the interval
    $(0,\delta)$ for sufficiently large $s$, where $\delta$ is arbitrarily
small. Since the sum of the infima does not exceed the infimum of the sum,
we conclude that (\ref{lab42}) implies (\ref{lab31}). By Theorem 1, the
series $\sum_KD_K\,W^K$ is unconditionally summable in $S^{\prime b}_a$,
and the number series (\ref{lab18}) is absolutely summable for any test
function $f\in S^b_a$. To complete the proof, it remains to apply Lemmas 1
and~2.

   {\bf Remark.} The temperedness of growth means that the singularities of
   $w_{{\rm maj}}$ are no worse than polynomial or logarithmic
ones. Therefore, it can be assumed that the inequalities
$$
   w_{{\scriptscriptstyle
   IR}}(\lambda r)\leq C_\lambda w_{{\scriptscriptstyle  IR}}(r), \qquad
      w_{{\scriptscriptstyle UV}}(t/\lambda )\leq C'_\lambda
      w_{{\scriptscriptstyle UV}}(t),
$$
hold for any $\lambda>0$, at least in the limiting sense, i.e., for
$r>R(\lambda)$ and $t<\delta(\lambda)$. In this case, the criterion
(\ref{lab42}) takes a simpler form
\begin{equation} \sum_k
L^kk!d_{2k} w_{{\scriptscriptstyle IR}}(r)^k \le C_{L}\,a(r),\quad
    \inf_{t>0}\,e^{st} \sum_k L^kk!  d_{2k} w_{{\scriptscriptstyle
      UV}}(t)^k \le C_{L}\, b(s),
\label{lab48}
\end{equation}
where $L>0$ is arbitrarily large. In particular, the conditions
(\ref{lab48}) when applied to the normal exponential $:\exp ig\phi:(x)$
become
$$
\exp\{L\,w_{{\scriptscriptstyle IR}}(r)\} \le
      C_{L}\,a(r), \quad \inf_{t>0}\,\exp\{st+L\,
      w_{{\scriptscriptstyle UV}}(t)\} \le C_{L}\, b(s).
$$

      \section{The case of a positive metric}
We now apply the suggested construction to the simplest case of a free
scalar field with the Wightman function
$$
\Delta^+(x;m)={1\over (2\pi)^{{\tt d}-1}}
\int \theta(p^0) \delta (p^2-m^2) e^{-ipx}{\rm d} p ={1\over (2\pi)^{{\tt d}-1}}
\int {e^{-i\omega({\bf p})x^0}\over 2\omega({\bf p})}e^{i{\bf px}}{\rm d}{\bf
p}.
$$
Let the dimension of the space-time be ${\tt d}>2$. Using the inequality
$\omega({\bf p})=\sqrt{{\bf p}^2+m^2}\geq\varepsilon|{\bf p}|+
m\sqrt{1-\varepsilon^2}$ and the notation $s=|{\bf p}|$,
    $m'=m\sqrt{1-\varepsilon^2}$, we derive the elementary estimate
$$
|\Delta^+(x^0-it,{\bf x};m)|\leq
   Ce^{-m't}\int_0^\infty e^{-\varepsilon st}s^{{\tt d}-3}{\rm d}
   s=C'e^{-m't}/t^{({\tt d}-2)}
$$
showing that
$w_{{\scriptscriptstyle  IR}}=0$ and $w_{{\scriptscriptstyle
UV}}(t)=e^{-m't}/t^{({\tt d}-2)}$ can be taken in this case.
In view of the infrared boundedness of the distribution $\Delta^+$, we can
take $S^0$ as the initial test function space for the Wick series. For the
same reason, Wick series can be convergent even when the
localization condition (\ref{lab40}) is violated. Theorem 2
permits analyzing this situation and strengthening Rieckers'
results \cite{Rieckers} (see the following theorem).

   {\bf Theorem 4}. {\it Let $\phi$ be a free neutral scalar field of mass
   $m>0$ in a space-time of dimension ${\tt d}>2$. The Wick series
$\sum_{k=0}^\infty d_k:\phi^k:(x)$ with positive
coefficients satisfying the condition $(\ref{lab41})$ is well defined
in the Fock space of the field $\phi$ under smoothing with test
functions in the space $S^b$ for which
\begin{equation}
\sum_{k<s} k!d_{2k}(s/k)^{k({\tt d}-2)}\leq C_\epsilon b(\epsilon s)
\label{lab50}
\end{equation}
with an arbitrary $\epsilon>0$. If $m= 0$, then this series is well
defined under the stronger condition
\begin{equation}
\sum_{k=0}^\infty k!d_{2k}(s/k)^{k({\tt d}-2)}\leq C_\epsilon b(\epsilon s).
\label{lab51}
\end{equation} }

{\bf Proof.} The theorem can be proved by repeating the proof
of Theorem 3 with obvious changes of some details.
The main distinction is that in Theorem 3 we first estimated the sum
of analytic functions and only then took the infimum over $t$, whereas
here we choose a free parameter in the representation
(\ref{lab30}) for each individual term of the series.
Using the same notation, applying (\ref{lab43}) and (\ref{lab44}),
and setting $N=2n{\tt d}+1$, we obtain
   \begin{equation}
   \sum_K \inf_{t>0} e^{st}
   \int\frac{|{\bf v}_K(x+it\eta)|}{(1+|x|)^N}\, {\rm d}x\leq
   C\,\sum_{k=0}^\infty L^kk!d_{2k}\inf_{t>0}
   \frac{e^{(s-km')t}}{t^{k({\tt d}-2)}}.
   \label{lab52}
   \end{equation}
Let $m\ne 0$. We write $\lambda=e/({\tt
     d}-2)$.  The infimum in (\ref{lab52}) is equal to zero for $k\geq
s/m'$ and is equal to $(\lambda(s/k-m'))^{k({\tt d}-2)}\leq (\lambda
     s/k)^{k({\tt d}-2)}$ for other values of $k$. We assume that
     $\lambda'=L^{1/({\tt d}-2)}\lambda\geq 1/m'$. It then follows from
     (\ref{lab50}) that the right-hand side of (\ref{lab52}) is majorized
     by $C_\epsilon b(\epsilon \lambda's)$, which ensures the fulfillment
     of the conditions of the Theorem 2 because $\epsilon$ is arbitrary.
     For $m=0$, no truncation of summation occurs, and the same result is
     ensured by the formula (\ref{lab51}). Applying Lemmas 1 and 2
     completes the proof.

     We note that in the localizable case, the variation of the plane of
     integration in each individual term of the series gives no significant
     improvement of the estimate and the conditions (\ref{lab50}) and
     (\ref{lab51}) become equivalent.  Indeed, let the coefficients
     $d_k$ have the form
 \begin{equation}
 d_k=k!^{-1/\rho}, \quad \rho>0.
 \label{lab53}
 \end{equation}
 Simple calculations using the Stirling formula that are similar to
 those in Sec.IV.2.2 in \cite{GelfandShilov} show that the functions
$$
\inf_{t>0}\,e^{st}\sum_{k=0}^\infty k!d_{2k}/t^{k({\tt d}-2)}
$$
and
$$
\sum_{k=0}^\infty k!d_{2k}\inf_{t>0}\,e^{st}/t^{k({\tt d}-2)}=
\sum_{k=0}^\infty k!d_{2k}(es/k({\tt d}-2))^{k({\tt d}-2)}
$$
have the same order of growth $\varrho=({\tt d}-2)/({\tt d}-3+2/\rho)$
for $\rho<2$ and differ only in their types. The maximum term in the second
of these series corresponds to the index $k(s)\sim s^\varrho$, and the
restriction of the summation to the range $k<s$ therefore does
 not affect the asymptotic behavior for $\varrho<1$, which is equivalent to
$\rho<2$. In contrast, for $\rho>2$, the truncated series behaves as
 $e^{(1-2/\rho)s\ln s}$ as $s\to \infty$. In particular, all entire
functions of a massive field with a finite order of growth are well defined
 on any of the spaces $S^\beta$ with the index $\beta<1$ and even on
 $S^b$, where $b(s)=s(\ln(1+s))^\gamma$, $\gamma>1$, whereas the
 universal space for realizing finite-order entire functions in the
 massless case is the narrower space $S^{1-1/({\tt d}-2)}$.

       \section{Wick series for the two-dimensional massless field}

 We consider the simplest example of a field with singular infrared
 behavior. Quantizing such a field requires introducing an indefinite
 metric (see \cite{Wightman}). Pierotti constructed the normal exponential
 of this field with an exact description of the test function space
 \cite{Pierotti}. Here, we describe suitable test functions for realizing
 the Wick entire functions of this field that have an exponential growth
 with the order $\rho<2$. For ${\tt d}=2$, the singular expression
 $2\pi\theta(p^0) \delta (p^2)$ has a clear meaning on the subspace
 $S_\bullet\subset S$ consisting of the functions vanishing at $p=0$.
 It is written as $\pi(\theta(u)u^{-1}\delta(v)+\theta(v)v^{-1}\delta(u))$
 in the light cone variables $u=p_0+p_1$ and $v=p_0-p_1$. Its Lorentz
 invariant extension to $S$ has the form
\begin{equation}
 \pi(u_+^{-1}\delta (v) + v_+^{-1}\delta(u)) +
 c\,\delta (u) \delta (v),
\label{lab54}
\end{equation}
where
$$
  u_+^{-1}(f) \stackrel{{\rm def}}{=} -\int_0^\infty f^{\prime}(u) \ln u\,
  {\rm d}u,
$$
and is not positively definite for any value of the constant $c$.
Passing to Laplace transforms, we obtain
$$
 w(z)=-\frac{1}{4\pi}\ln (-\kappa^2z^2)\qquad ({\rm Im}\,z\in \oV_-),
$$
where $\kappa$ is an arbitrary scaling parameter whose exact relation to
the constant $c$ in (\ref{lab54}) is insignificant. Let $h\in S$ and $\hat
h(0)=1$. Then $\hat f_0=\hat f-\hat f(0)\hat h\in S_\bullet$. We write
$$
\langle f,g\rangle=\int \bar f(x)w(x-x')g(x)\,{\rm d}x{\rm d}x'
$$
and set $\langle h,h\rangle=0$, which can be ensured by an appropriate
scale transformation. In this case, we have
 $$
 \langle f,g\rangle=\langle f_0,g_0\rangle +\,\,\bar{\!\!\hat f}(0)\langle
 h,g\rangle +\hat g(0)\langle f,h\rangle.
 $$
According to \cite{Pierotti, MorchioStrocchi}, $w_{{\rm maj}}$ is given by
the relation
 $$
 \int \bar f(x)w_{{\rm maj}}(x,x')g(x')\, {\rm d}x {\rm
 d}x'= \langle f_0,g_0\rangle+\langle f,h\rangle\langle h,g\rangle +
 \,\,\bar{\!\!\hat f}(0)\hat g(0).
 $$
If $h$ is a real-valued even function, then the explicit form of the
majorant is
   $$
 w_{{\rm  maj}} (x,x')=w(x-x')+ (w_h(x)-1)(w_h(-x)-1),
   $$
where $w_h=w*h$. Let ${\rm Im}\,z=-t e_0$,  $e_0=(1,0)$ and $t>0$. Then
$t^2\leq |z^2|$. On the other hand, $|z^2|\leq 2|z|^2$, and consequently
 $|\ln(-z^2)|\leq C+2\ln(1+|z|)+2\ln^+(1/t)$, where
 $\ln^+(\cdot)=\max(0,\ln(\cdot))$. Let $h\in S^0$. Then $w_h$ is
 an entire function satisfying the inequality
 $|w_h(z)|\leq C+ C'\ln(1+|z|)$ for the indicated values of ${\rm Im}\,z$.
 It can be easily established using the representation
 $$
 w_h(z)=\int h(\xi+ie_0)w(z-\xi-ie){\rm  d}\xi
 $$
 and the inequality $1+|z-\xi|\leq (1+|z|)(1+|\xi|)$.
 We can therefore take
 $$
 w_{{\scriptscriptstyle  IR}}(r)= (\ln  (1+r))^2,\qquad
 w_{{\scriptscriptstyle UV}}(t)=\ln^+(1/t).
 $$
 in the case under consideration. In this case, the role of the test
 function space for realizing any Wick entire function of order $\rho<2$
 can clearly be played by $S^\beta_\alpha$ with arbitrarily large
 $\alpha$ and $\beta$. It is also easy to obtain a more exact description.
 For the coefficients $d_k$ of the form (\ref{lab53}), the inequalities
$$
   \sum_k
   L^kk!d_{2k} w_{{\scriptscriptstyle IR}}(r)^k \le Ce^{N(\ln
  (1+r))^{2\rho/(2-\rho)}},\,\, \inf_{t>0}\,e^{st} \sum_k L^kk!  d_{2k}
   w_{{\scriptscriptstyle UV}}(t)^k \le Ce^{N(\ln (1+s))^{\rho/(2-\rho)}}
$$
 hold, where $N$ depends on $L$ and the infimum can be estimated by
 the value of the function at the point $t=1/s$, which does not in fact
 lead to cruder results. In particular, if $\rho\leq 2/3$,
 then the conditions (\ref{lab42}) hold for $a(r)=(1+r)^N$ and $b(s)=1+s$.
 Hence, for this order of growth, the operator realization with test
functions in the Schwartz space $S$ is possible, which can be seen if
$S^0$ is taken as the original test function space and an analogue of
Lemma 3 for the simplest case of the extension of $u\in S'_0$ to $S$
is used. Namely, such an extension is sure to exist,
if $|(u*g)(p)|\le C_B\|g\|_{B,N}\,(1+|p|)^N$ for some $B$ and $N$.
For $2/3<\rho\leq 1$, the behavior is polynomially bounded with respect to
  $s$ as before but is no longer polynomially bounded with respect to $r$,
  namely it is characterized in the latter case by the function
  $e^{(\ln(1+r))^\alpha}$, where $\alpha=2\rho/(2-\rho)$.  If $\alpha<2$
  (i.e., $\rho<1$), then the function with such a slow growth  cannot
  serve as an indicator function because it does not satisfy the condition
  (\ref{lab23}). The space $S_a$, where $a=e^{(\ln(1+r))^2}$, is already
  acceptable and can serve as the test function space for any $\rho<1$ by
  Theorem 2.  For $\rho=1$ and a finite order of growth, the series
  (\ref{lab18}) converges on the functions in $S_{a_N}$, where
  $a_N=e^{N(\ln(1+r))^2}$ and $N$ grows with increasing $n$. Therefore,
  the adequate space for this case is ${\cal P}_2=\bigcap_N S_{a_N}$.
  It consists of smooth functions such that the norms
  $$
  \|f\|_N=\sup_x\max_{|\kappa|\leq
 N}|\partial^{\,\kappa}f(x)|e^{N(\ln(1+r))^2} \qquad (N=0,1,2,\ldots).
    $$
  are finite and belongs to the class $K(M_N)$ (see \cite{GelfandShilov}
  for the theory of these spaces). For $\rho>1$, it is natural to use
  the spaces
  $$
  {\cal P}^\beta_\alpha=\bigcap_{N=0}^\infty
  S_{a_{\alpha,N}}^{b_{\beta,N}},
  $$
  where
  $$
  a_{\alpha,N}(r)=e^{N(\ln(1+r))^\alpha},\quad
  b_{\beta,N}(s)=e^{N(\ln(1+s))^\beta},\qquad \alpha, \beta\geq 2.
  $$
  Ultimately, we obtain the following result.

   {\bf Theorem 5.} {\it  Let $\phi$ be a free massless scalar field in
  a space-time of dimension $2$. The role of the test function space for
  its normally ordered entire functions of the order $\rho<2$
  and of finite type can be played by the following spaces:

   the Schwartz space $S$ for $\rho\leq2/3$;

   the space ${\cal P}_2$ for all functions of the exponential type;

   the space ${\cal P}_{2\rho/(2-\rho)}^2$ for $1<\rho\leq 4/3$;

   the space ${\cal P}_{2\rho/(2-\rho)}^{\rho/(2-\rho)}$ for
   $4/3<\rho< 2$.}

       \section{Conclusion}

The presented theorems reduce finding the test function space for normally
ordered power series of a free field to estimating the order of growth
and the order of singularity for the Hilbert majorant of the two-point
  function of the field. If the adequate test function space contains
  functions of compact support in both the coordinate and momentum
  representations, then the limit of the Wick series satisfies the main
  general requirements of quantum field theory
  \cite{BLOT} in a practically obvious manner because we have
  constructed the operator realization in the very state space
  of the original free field. However, as a rule, the
  localizability condition does not hold in momentum space for
  the exact solutions of gauge models in a generic covariant
  gauge \cite{MoschStroc, Moschella}, and the Wick series involved in them
  are defined only on test functions whose Fourier transforms are
  analytic. The statement of the spectral condition and the verification
  of its fulfillment is a nontrivial problem in this case
  \cite{Moschella,Soloviev3}. Moreover, this problem is of particular
  interest because its solution is instructive for the correct
  generalization of the reconstruction theorem and the Osterwalder-Schrader
  Euclidean theory to quantum field theory with infrared singular
  indefinite metric, where the Poincar\'e group is implemented by
  pseudounitary operators. We plan to investigate this circle of problems
  in a separate publication. Another interesting problem whose solution can
  be obtained by applying the approach developed in this paper is the problem
  of
  constructing the nonlocal extension of the Borchers equivalence classes
  and proving that the nonlocal Wick series of a free field with
  a positive metric satisfy the condition of asymptotic commutativity
  \cite{Soloviev4} which ensures the preservation of the main physical
  consequences in nonlocal quantum field theory.

  {\bf Acknowledgments.} This work is supported by the Russian Foundation
  for Basic Research (Grant No. 99-01-00376) and INTAS (Grant No. 96-0308).

               \end{document}